# Mode Modification of Plasmonic Gap Resonances induced by Strong Coupling with Molecular Excitons

Xingxing Chen[1†], Yu-Hui Chen[2†], Jian Qin[1], Ding Zhao[1], Boyang Ding[2]*
Richard J. Blaikie[2*], and Min Qiu[1]*

[1]*State Key Laboratory of Modern Optical Instrumentation, Department of Optical Engineering, Zhejiang University, Hangzhou 310027, China*
[2] *MacDiarmid Institute for Advanced Materials and Nanotechnology, Department of Physics, University of Otago, PO Box 56, Dunedin 9016, New Zealand*

\* boyang.ding@otago.ac.nz
\* richard.blaikie@otago.ac.nz
\* minqiu@zju.edu.cn

**Abstract:** Plasmonic cavities can be used to control the atom-photon coupling process at the nanoscale, since they provide ultrahigh density of optical states in an exceptionally small mode volume. Here we demonstrate strong coupling between molecular excitons and plasmonic resonances (so-called plexcitonic coupling) in a film-coupled nanocube cavity, which can induce profound and significant spectral and spatial modifications to the plasmonic gap modes. Within the spectral span of a single gap mode in the nanotube-film cavity with a 3-nm wide gap, the introduction of narrow-band J-aggregate dye molecules not only enables an anti-crossing behavior in the spectral response, but also splits the single spatial mode into two distinct modes that are easily identified by their far-field scattering profiles. Simulation results confirm the experimental findings and the sensitivity of the plexcitonic coupling is explored using digital control of the gap spacing. Our work opens up a new perspective to study the strong coupling process, greatly extending the functionality of nanophotonic systems, with the potential to be applied in cavity quantum electrodynamic systems.

---

[†] These authors contributed equally to this work

Strong light-matter interactions build upon fast energy exchange [1] between excitonic quantum emitters and electromagnetic (EM) modes in a cavity, which leads to cavity-exciton mode hybridization, manifesting as a split in the cavity spectral response, known as vacuum Rabi splitting. Understanding these phenomena is very important for cavity quantum electrodynamics applications [2,3]; for example, quantum computing requires repeated manipulation of excitonic states with photons before atomic or molecular excitons lose their coherence. On the other hand, once strongly coupled to a resonant cavity, electronic states of molecules can be greatly reshaped [4] from those in free space, enabling a variety of extraordinary effects, such as modification of molecular thermodynamics [5] and chemical landscape [6], mediation of energy transfer between multiple molecules [7] and even alternation of photosynthetic processes in bacteria [8].

Strong coupling (SC) requires a high atomic cooperativity ($C = g^2 / \gamma \cdot \kappa$) [9], where $g$ is the coupling strength, $\gamma$ and $\kappa$ are the spectral width of the excitonic transition of emitters and EM modes respectively. To achieve high $C$, traditional investigations [9–12] have used cavities with high $Q$ performance (quality factor $Q = \omega / \kappa$, where $\omega$ is the oscillation frequency of the cavity). Another solution is to reduce the cavity mode volume $V$, since $g \propto \sqrt{N/V}$, where $N$ is the number of excitons contributing to the coupling process. In this context, plasmonic cavities have been proposed to investigate strong coupling, because these cavities can concentrate light energy within deep sub-wavelength volumes, and thus are capable of providing high cooperativity by compensating their moderate $Q$ values with ultra-small $V$ [13–15]. Specifically, plasmonic cavities are specially designed metallic nanostructures, in which collective oscillation of conduction electrons, known as localized surface plasmons, can be excited upon illumination. The interaction between plasmonic resonances and nearby atomic or molecular excitons gives rise to so called plexcitonic coupling [16].

For convenience of measurements, previous experiments exploring plexcitonic coupling were primarily based on ensemble metallic nanostructures, including arrays of nanovoids [17,18] or nanoparticles (NPs) [19–23] and nanocrystals in solution [16,24–26]. Recently several groups have moved to a much smaller scale, investigating strong interaction of molecular excitons with individual plasmonic resonators, such as metallic nanospheres [27,28], nanorods [29–31], nanoprisms [15] and nanodisk dimers [13]. This greatly enlarges the freedom to observe and control the strong coupling process at the nanoscale, significantly deepening our understanding of relevant phenomena. Notably, two recent studies have reported the strong coupling of a single quantum dot with a gold (Au) NP [14] and bowtie structures [32], indicating that the nanoscale dimensions make plasmonic resonators a very promising platform for realizing light-matter interactions at the fundamental limit of single atom-photon coupling, which is crucial for potential quantum optics applications [2,3,9].

In this letter we report strong coupling in plasmonic cavities with highly compressed EM mode volumes, i.e. single Au nanocube-film (NC-film) resonators (Figure 1a) that can squeeze EM fields into an extremely narrow gap (a few nanometers wide, typically less than $\lambda/100$) between nanoparticles and a metallic substrate [33–37]. The exceptionally small $V$ significantly enhances the density of optical states, even enabling single molecules strong coupling in the cavity [38]. Here using the NC-film resonators, we for the first time experimentally demonstrate that by doping the nano-gap with J-aggregate dye molecules, the greatly enhanced strong coupling between molecular excitons and gap plasmons not only enables the spectral Rabi splitting, but also, more importantly, shows significant modifications of far-field scattering patterns from the NC-film cavity. Comparing the experimental results with numerical simulations reveals that the scattering pattern modification is induced by the remarkably altered spatial profile of the cavity mode, i.e. the underlying and well-understood single spatial gap mode transforms to take on two distinctive and significantly different spatial profiles above and below the plexcitonic coupling wavelength.

Similar phenomena have been theoretically predicted [39], where De Liberato notes that if the light-matter interaction is strong enough, the spatial profile of cavity modes can be significantly altered, surprisingly leading to the decoupling between light and matter, i.e. the breakdown of Purcell effect, while the spontaneous emission rate of the coupled quantum emitter dramatically decreases. This predicted effect is of great importance for applications of plasmonic cavities, as plasmonic enhancement of light-matter interaction has been extensively utilized to increase spontaneous emission rate in various plasmonic structures. However there is little knowledge about how the strong coupling induced mode modification affects plasmonic enhancement. In this context, our study not only provides insights into fundamental understandings of coherent strong coupling, but also gives a preliminary demonstration of spatial mode modification induced by plexcitonic coupling, which may greatly advance the investigation of plasmonic enhancement from the perspective of light-matter interaction. In addition, our study also shows alternative measures to effectively improve the functionality of nanophotonic structures, e.g. reshaping radiative properties of plasmonic resonators by inducing quantum emitters but not merely modifying physical structures [44].

In particular, as shown by the schematic (Fig. 1a) and the scanning electron microscope (SEM) image (inset of Fig. 1a), AuNCs acquiring an averaging side length of $65 \pm 5$ nm were separated from a Au film by a dielectric spacer that is self-assembled using polyelectrolyte (PE) layers, and then coated with a dilute layer of J-aggregates. Applying different number of PE layers allows exquisite and precise control of the gap spacing $d$ between AuNCs and the metallic substrate. Fig. 1(b) illustrates the optical characterisation setup for the NC-film cavity, where a dark-field objective is used to collect scattered light from single NCs dispersed on substrates. Details of experiments and modelling can be found in Section 1 of the Supplementary Information.

Figure 1(c) shows a typical scattering spectrum (black curve) measured from dye-free NC-film cavities with 3 PE layers, which provides a gap spacing of $d = 4.3 \pm 0.7$ nm between NCs and the film. Such a gap spacing gives rise to a scattering maximum at 640 nm (1.94 eV) as a result of the excitation of a gap resonance [34,35]. In contrast, when the nano-gap is doped with J-aggregate molecules that possess overlapped absorption with the gap resonance spectrum (as shown by the blue dashed curve in Fig. 1c), the scattering properties of the NC-film cavity are significantly modified. Specifically, the doped NC-film cavity (Fig. 1d) shows two pronounced split scattering maxima at 623 nm and 684 nm respectively with a deep minimum at 649 nm that shares the same spectral position as the molecular absorption peak, signifying that the plasmonic modes are strongly coupled with molecular excitons [13–15,27–32]. With gap spacing set to be $d$ = 3 nm, the simulated spectra for both undoped (Fig. 1e) and doped (Fig. 1f) cavities agree very well with measured results.

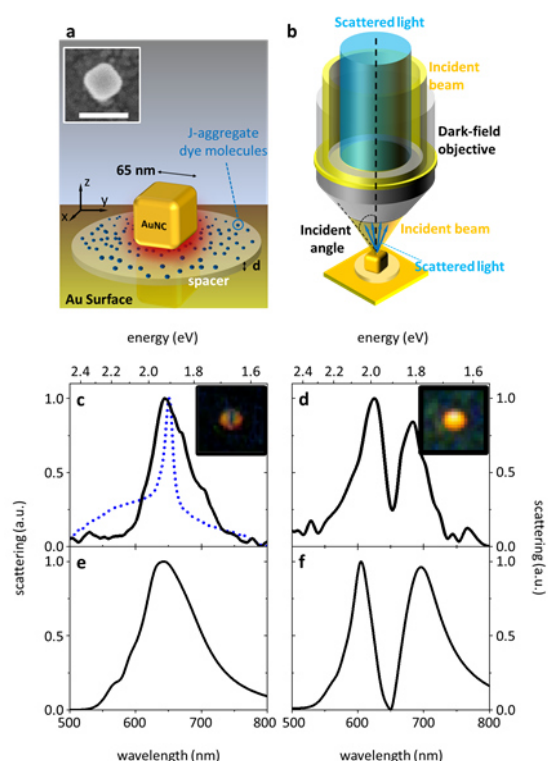

**Figure 1** (a) A schematic of a NC-film cavity doped with J-aggregate dye molecules, where the red color represents light confined in the gap. A top-viewed SEM image of the NC-film cavity is shown as an inset with a white bar standing for 100 nm. (b) A schematic of a dark-field microscope to measure scattering from a sample. Measured (c/d) and simulated (e/f) scattering spectra of the undoped/doped NC-film cavity for the gap spacing (3 PE layers) with insets: the corresponding far-field scattering images. The absorption spectrum (blue dashed curve) of J-aggregate dye molecules in water is shown in panel (c).

Very noteworthy here is that the spatial scattering profiles are also significantly altered. Specifically, the far-field scattering image of the undoped cavity (inset in Fig. 1c) shows a doughnut-shaped pattern, a typical feature where the electric fields in the gap are vertically excited with respect to the film surface [34]. In contrast, the doped NC-film cavity exhibits a dot-like scattering pattern (inset in Fig. 1e). We note that this is a systematic effect and not an artifact due to subtle variations in nanocube geometry or gap spacing, as scattering images containing a number of NCs in the same field of view all show identical dot-like scattering patterns (Fig. S2c and Fig. S4a in the Supplementary Information). The dot-like scattering patterns of 3 PE layer doped cavities are dramatically different from those of undoped cavities with small gap spacing, e.g. 2, 3 and 4 PE layer spacer (doughnut patterns as shown in Fig. S2a, S2b and S2c in the Supplementary Information), while share some shape similarities with larger spacing undoped cavities (5 or 7 PE layer spacer), so one possible reason could be thickness increase induced by the additional coating of J-aggregates. However thickness measurements reveal that the J-aggregates coated 3 PE layer spacer has a thickness of $d = 3.9 \pm 0.5$ nm, indicating that the added dye layer doesn't change the gap spacing significantly. Given that such dye molecules usually acquire a size of ~1 nm [45], the unchanged gap spacing is likely due to the de-swelling of PE spacers [46,47] after being immersed in the acidic J-aggregate solution followed by being rinsed with de-ionized water in the fabrication process (detailed discussions can be found in the Section 1 and Section 3 of the Supplementary Information).

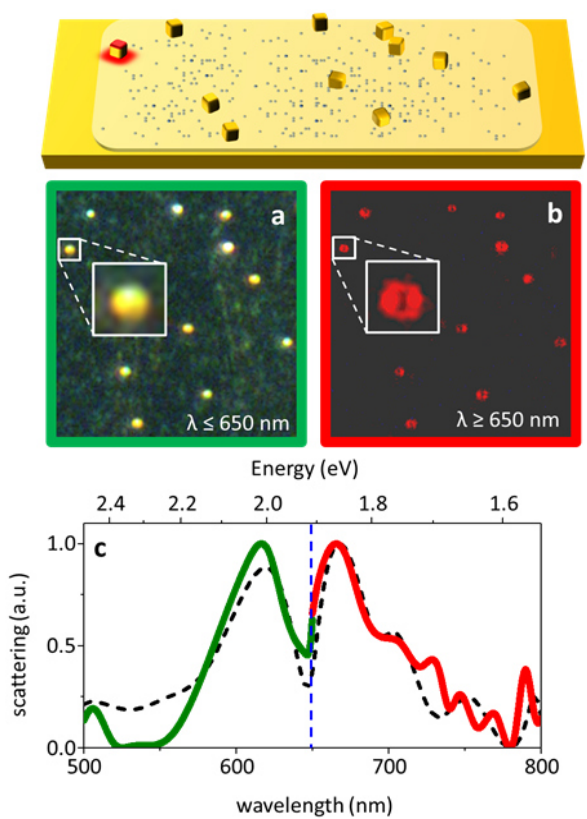

**Figure 2** Far-field scattering images taken from the same area of a sample (the doped NC-film structures with gap spacing d = 3.9±0.5 nm) measured with a short-pass filter (a) λ ≤ 650 nm and a long-pass filter (b) λ ≥ 650 nm. (c) The scattering spectra of the NC labeled with white frames and magnified in panel (a) and (b) measured using a filter λ ≤ 650 nm (green curve), a filter λ ≥ 650 nm (red curve) and no filter (black dashed curve), while the blue dashed line is to indicate spectral position of molecular excitons.

To reveal the physical nature of the scattering pattern modification in our NC-film cavity system, we have performed filter-based scattering measurements. In particular, a dark-field microscope integrated with a short- (long-) pass filter was used to take scattering images on the same area of a Au substrate covered with a J-aggregate-coated PE spacer ($d = 3.9 \pm 0.5$ nm) and NCs. The cut-off (-on) wavelength of the short- (long-) pass filter is at $\lambda = 650$ nm, allowing us to selectively observe and analyze the optical properties of each individual component of the split maxima in the scattering spectra (Fig. 2c). Specifically, the scattering patterns of NCs taken at the short wavelength band ($\lambda \leq 650$ nm) all display a dot-like shape (Fig. 2a), whereas most of the same particles exhibit doughnut-like patterns at the long band ($\lambda \geq 650$ nm) (Fig. 2b), similar to what we observed in undoped NC-film cavities (inset in Fig. 1c). Far-field patterns directly relate to the near-field distributions in a cavity, while the filter-based measurements clearly demonstrates that the two split maxima acquire different spatial modal characteristics from each other in the SC regime, even though they originate from a single broad scattering resonance in the undoped cavity system. This implies a complex interaction behavior between light and matter in the NC-film cavity.

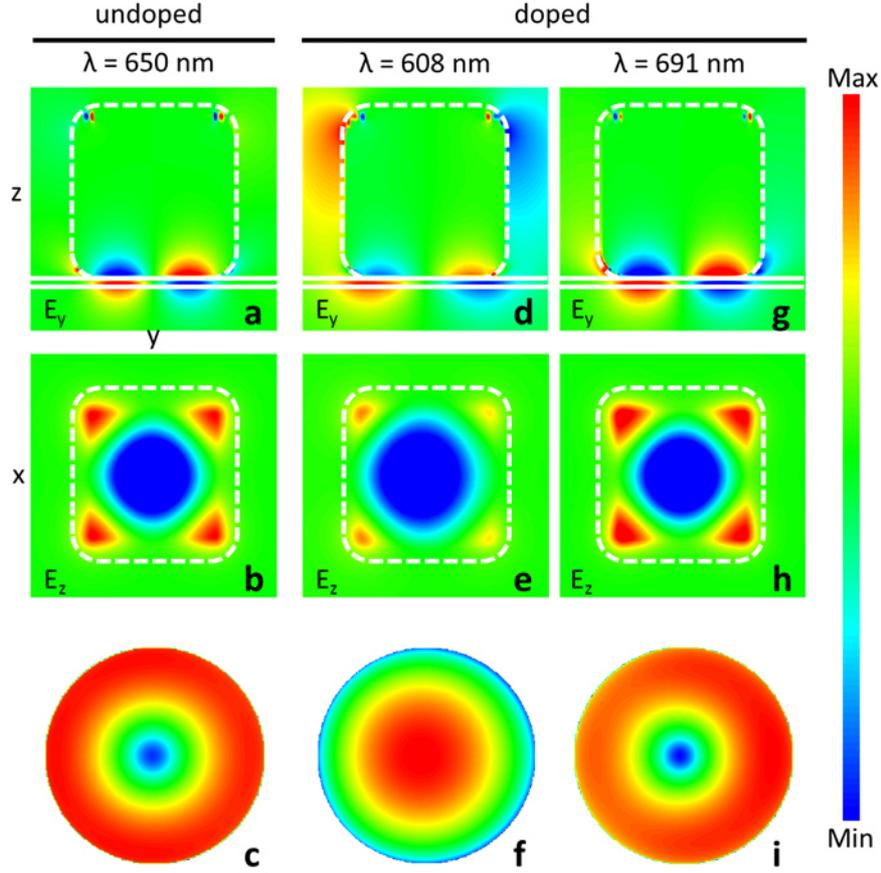

**Figure 3** (a) the y-component of electric fields ($E_y$, viewed from the y-z plane mid cross-section of the nanocube), (b) the z-component of electric fields ($E_z$, viewed from the x-y plane mid cross-section of the nano-gap) and the far-field scattering pattern (c) for the resonance in the simulated spectrum of the undoped NC-film cavity with d = 3 nm (black curve in Fig. 1e). (d/g) the y-component of electric fields, (e/h) the z-component of electric fields and the far-field scattering patterns (f/i) for the short/long wavelength maxima in the simulated spectrum of the doped NC-film cavity with d = 3 nm (black curve in Fig. 1f). Red (blue) areas in all panels correspond to positive (negative) maximum amplitude.

Numerical analyses provide more insights into these phenomena. Specifically, we have carried out simulations to demonstrate electric field distributions in undoped and doped NC-film cavities. Figures 3(a) and 3(b) show the electric field's y-component $E_y$ observed from the y-z plane mid cross-section of the NC-film cavity and z-component $E_z$ viewed from the x-y plane mid cross-section of the gap respectively, corresponding to the resonance at $\lambda=650$ nm in the modelled spectrum of the undoped cavity (Fig. 1e). It manifests that most of the electric fields are tightly confined within the gap (Fig. 3a) with field maxima at the corners and center of the square gap area (Fig. 3b), exhibiting a typical feature of out-of-plane plasmonic gap resonances in NC-film structures [41] as well as inducing a doughnut-like pattern in the simulated far-field scattering image (Fig. 3c), which is in a very good agreement with experimental result (inset in Fig. 1c). As discovered in experiments, doping the nano-gap with J-aggregates leads to a significant transition of the cavities' optical behaviors, which can also be observed in simulated near-field and far-field properties. For example, in the doped cavity, it is noted that at the spectral position of the short band maximum ($\lambda = 608$ nm) of the simulated scattering spectrum (Fig. 1f), the $E_y$ component (Fig. 3d) displays two lobes at the top corners with suppressed diffusion into the gap, while the $E_z$ component (Fig. 3e) shows hot-spots at four corners significantly diminish with the center fields spreading over the gap area. Such near-field distributions are very different from those in the undoped cavity (Fig. 3a and 3b), resulting in a dot-like far-field pattern (Fig. 3f). In contrast, at the long band split maximum ($\lambda = 691$ nm) it exhibits a doughnut shape far-field pattern (Fig. 3i), greatly resembling to those of the undoped cavity. It is interesting to note that such a resemblance can also be observed in the theoretical discussion about SC induced spatial mode modification [39].

These results compare favorably with the filter-based observations for the short (Fig. 2a) and long (Fig. 2b) band measurements of doped cavities, demonstrating distinct far-field patterns at different split maxima. This further confirms that the strong plexcitonic coupling in the NC-film cavity system not only induces spectral modifications as revealed in previous studies, but also dramatically alters the near-field spatial distribution of gap modes that directly relate to the variation of far-field scattering patterns in the experimental observations.

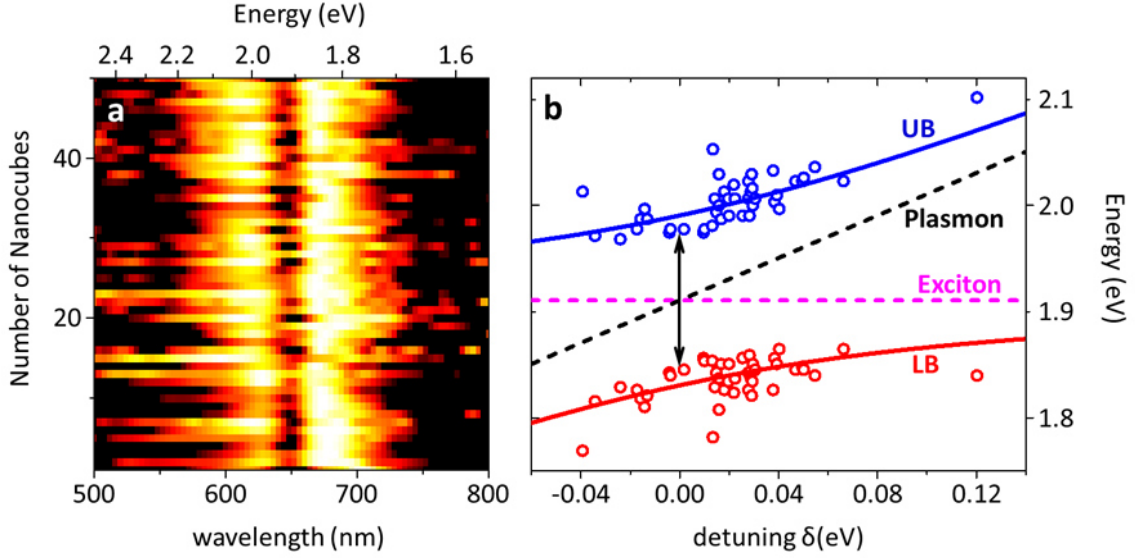

**Figure 4** (a) Scattering spectra of doped NC-film cavity with gap spacing d = 4.0±0.5 nm as a function of measurement counts. (b) Hybridized mode frequencies ($\omega_+$: blue open circles; $\omega_-$: red open circles; extracted from scattering maxima in panel **a**), cavity plasmon (black dashed line) and molecular exciton (magenta dashed line) plotted against the detuning δ. Calculated dispersion curves (blue and red lines) based on formula (1) in the text to fit the measured data. The plot resembles anti-crossing behavior with average vacuum Rabi splitting of $\Omega_R \approx$ 170 meV.

To characterize the plexcitonic coupling strength that can induce spatial modal transition, we have measured scattering spectra of over 50 doped NC-film cavities with gap spacing $d = 3.9 \pm 0.5$ nm. Although fluctuations in geometries of AuNCs, thickness of PE spacers and number of J-aggregates in each cavity may lead to variations in scattering properties, the spectral map in Fig. 4(a) demonstrates clear non-dispersive minima at ~649 nm flanked by two maxima for all measured cavities, unambiguously indicating these samples all show pronounced coupling effects. In order to determine the extent of strong plexcitonic coupling in our system, we have studied the split behavior using a coupled harmonic oscillator model, which can describe the plexcitonic coupling with the Jaynes-Cummings quantum mechanical model, [13,15]

$$\omega_{\pm} = \frac{1}{2}(\omega_{pl} + \omega_0) \pm \sqrt{g^2 + \frac{\delta^2}{4}} \qquad (1)$$

where $\omega_+$ and $\omega_-$ represent the upper (UB) and lower (LB) plexcitonic branches, $\omega_{pl}$ and $\omega_0$ are plasmon and excitonic energies respectively, and the detuning $\delta = \omega_{pl} - \omega_0$. Here we assume molecular excitonic energy is unchanged with $\hbar\omega_0 = 1.91$ eV, and then the coupling energy $\Omega_R$, also called Rabi splitting, can be given by $\Omega_R = 2\sqrt{(\omega_+ - \omega_0)(\omega_0 - \omega_-)}$ while the plasmon energy can be obtained by $\omega_{pl} = \omega_+ + \omega_- - \omega_0$. Using these equations, we plot the dispersion of split scattering maxima acquired from Fig. 4(a) as a function of the detuning $\delta$ in Fig. 4(b), using the coupling strength $g$ as a fitting parameter. It is evident to see the anti-crossing behavior of plasmonic and excitonic energies, where an average Rabi splitting $\hbar\Omega_R = \hbar \cdot 2g$ of ~170 meV is observed at the crossing point, which is far beyond the strong coupling criterion "$g/\omega_0 \geq 0.01$, $2g > (\kappa - \gamma)/2$ or $2g > (\kappa + \gamma)/2$" used in other works [13,48,49], where $\hbar\gamma = 59$ meV (the linewidth of excitonic transition) and $\hbar\kappa = 187$ meV (the spectral width of gap mode). As compared to the study [39] that theoretically discussed mode transitions in deep strong coupling regime ($g/\omega_0 \geq 1$), our result exhibits much lower coupling strength, manifesting that light-matter interaction induced spatial mode modification can also take place in a plasmonic resonator with exceptionally small mode volume.

In conclusion, we have studied the strong plexctonic coupling effects on NC-film cavities comprising of AuNCs placed on a metallic substrate with J-aggregate dye molecules embedded within the gap. As the result of strong interactions between gap modes and molecular excitons, scattering spectra of doped cavities show evidence of Rabi splitting features, and more importantly we discover that far-field scattering patterns of the NC-film cavity are significantly altered in SC regime, indicating that the hybrid plexcitonic modes acquire very different spatial characteristics from those of the uncoupled cavities. Numerical simulations are in a good agreement with experimental results.

These results are of particular interest to the study of plexcitonic coupling, since our work provides a preliminary experimental study that reveals the influence of strongly coupled excitons on the spatial mode characteristics of plasmonic cavities, which is a phenomenon that traditional investigations have usually neglected. These have enlarged our understanding of the physical nature of the light-matter interaction at the nanoscale and provided novel perspectives to understand plasmonic enhancement, thus are of great importance for relevant applications.

**Acknowledgements**

This work was financially supported by New Zealand's Marsden Fund through contract UOO-1214, the National Natural Science Foundation of China (Grants 61425023, 61575177, 61275030 and 61235007), the Priming Partnership Pilot Funding (University of Otago), the New Idea Research Funding 2016 (Dodd-Walls Centre for photonic and quantum technologies) and the Open Fund of China State Key Laboratory of Modern Optical Instrumentation.

**Author Contribution**

B.D. conceived the idea; X.C., J.Q., D.Z. and B.D. prepared the samples; X.C., J.Q., and B.D. carried out the optical and other characterisation; Y.H.C. and B.D. performed the simulation; M.Q., R.J.B. and B.D. supervised the projects; B.D. prepared the manuscript; all authors discussed and analysed the results.

# Mode Modification of Plasmonic Gap Resonances induced by Strong Coupling with Molecular Excitons

Xingxing Chen[1†], Yu-Hui Chen[2†], Jian Qin[1], Ding Zhao[1], Boyang Ding[2]*

Richard J. Blaikie[2]*, and Min Qiu[1]*

[1] State Key Laboratory of Modern Optical Instrumentation, Department of Optical Engineering, Zhejiang University, Hangzhou 310027, China

[2] MacDiarmid Institute for Advanced Materials and Nanotechnology, Department of Physics, University of Otago, PO Box 56, Dunedin 9016, New Zealand

* boyang.ding@otago.ac.nz

* richard.blaikie@otago.ac.nz

* minqiu@zju.edu.cn

# Supplementary Information

† These authors contributed equally to this work

**Section 1. Methods**

**Sample preparation**

AuNCs with an average side length 65±5 nm were prepared using a seed-mediated method [1,2] and confirmed by SEM imaging. The Au substrates were prepared by coating 100-120 nm thick Au films on smooth silicon wafers using electron beam evaporation. PE spacer layers were prepared using a digital layer-by-layer deposition method. Specifically, freshly prepared Au substrates were first immersed in poly(allylamine) hydrochloride (PAH) and sodium chloride (NaCl) aqueous solution (3 mM PAH and 1 M NaCl ) for 5 min, followed by being rinsed with deionized water, which leads to the formation of an approximately 1 nm thick PAH layer on the Au film. The PAH coated film was then immersed into polystyrene sulfonate (PSS) and NaCl aqueous solution (3 mM PSS and 1 M NaCl) for 5 min before being rinsed with deionized water again, resulting in a ~1 nm thick PSS layer attached to the PAH surface. This is because PAH molecules dissolved in water are positively charged while PSS are negatively charged. Alternating this procedure with PAH and PSS deposition allows us to produce a PE film on Au substrate with precisely controlled thickness determined by the number of deposition cycles.

The Au substrate coated with PE film was then immersed in the dye solution for 10 min followed by being rinsed with de-ionized water and then air dried to produce a dilute layer of J-aggregates on the top. The specific J-aggregates we use here were self-assembled from aqueous solution of the cyanine dye molecules 5-Chloro-2-[3-[5-chloro-3-(4-sulfobutyl)-3H-benzothiazol-2-ylidene]-propenyl]-3-(4-sulfobutyl)-benzothiazol-3-ium hydroxide, inner salt, triethylammonium salt, with a concentration of 1 mM. NaCl (0.1 M) was added to facilitate the aggregation of dye molecules in solution. These dyes were purchased from Few Chemicals GmbH (Product No. 2275). Adding hydrochloric acid (HCl with 1%wt) into the dye solution can facilitate the formation of J-aggregate layer on PE spacers. In addition, since the cyanine dye molecules used in our experiments are negatively charged if dissolved in water, the J-aggregates layer can only be attached to the positively charged PE surface. Therefore, for all doped NC-film cavities, their PE spacers must be terminated with PAH layer to adhere J-aggregate molecules, which leads to the odd number of the PE layers in these cavities, such as 3, 5 and 7 layers. Together with the structural formula of the cyanine dyes, a schematic of a J-aggregate embedded NC-film cavity with a 3-layer PE spacer is shown by the Fig S3A in the Supplementary Information, where we have also discussed PE spacers with additional thickness induced by J-aggregate coatings.

The thicknesses of all PE spacers and J-aggregate coated PE spacers were confirmed by ellipsometry measurements (GES 5E Semilab). It turns out that (i) the PE thickness is not linearly changed with the number of layers, which is likely the result of environmental humidity [3,4]; (ii) the addition of J-aggregate layer doesn't increase the thickness of PE spacer. Instead, in the case of 3 and 5 PE layers, the J-aggregate coated PE spacer acquired even thinner thickness than uncoated ones. This is likely due to the pH-dependent thickness of the PE spacer [5], while the J-aggregate solution shows a pH value of $0.7 \pm 0.3$ as a result of HCl mixing. The pH value was measured using a laboratory pH-meter (PHSJ-4A, Ningbo Hinotek Technology Co., Ltd.). More details about the thickness variation of PE spacer before and after being immersed in the J-aggregate solution can be found in the Section 3.3 of Supplementary Information.

In the last step, we deposited AuNCs on the PE-coated (or terminated with J-aggregates) substrates. Electrostatic deposition was considered to place AuNCs, but using this method doesn't allow efficient deposition of NCs on PAH-terminated spacer, as both PAH and NC surfaces are positively charged. To solve this, the AuNCs dispersed in aqueous suspension were drop-cast on the PE-coated substrates and allowed to dry. Before casting, the AuNC suspension was centrifuged several times to reduce impurities, which may lead to insufficient cleanliness in the optical measurement. From the scattering images over a large area of all samples (Fig. S2 and Fig. S4 in the Supplementary Information), we can see that nanoparticles can be clearly resolved using a dark-field microscope.

**Optical Characterisation**

We used a reflection dark-field microscope (Nikon, ECLIPSE 80) to measure the scattering spectra of individual AuNCs and to take ensemble images at lower magnifications. As shown in Fig. 1b, a dark-field objective (Nikon, CFI LU Plan Epi ELWD, 100X, NA = 0.8) was used to focus white-light illumination on the samples, and scattered light was collected by the same objective. A short-(long-) pass filter (FES650 and FEL650 purchased from Thorlabs Inc.) was placed after the objective to analyse specific resonances in scattering spectra for samples exhibiting strong plexcitonic couplings. The scattering spectra were recorded using a fibre-based spectrometer (Ocean Optics, QE65 pro), while scattering image were taken by a CCD camera (Nikon, DIGITAL CAMERA HEAD DS-Fi1).

**Simulations**

The spectra, near-field and far-field distributions of the NC–film system were simulated using Finite-Difference-Time-Domain (FDTD) method implemented by Lumerical Solutions. In simulations, the effective dielectric function of PE film coated with J-aggregates was calculated using a classical Lorentz model [6,7], assuming the PE layer homogeneously doped with J-aggregates:

$$\varepsilon_{\text{PE+Jagg}} = \varepsilon_{\text{PE}} + \Delta\varepsilon_{\text{Jagg}} \frac{\omega_{\text{abs}}^2}{\omega_{\text{abs}}^2 - \omega^2 - i\omega\gamma}$$

where $\varepsilon_{\text{PE}} = 2.07$ is the dielectric constant of PE films, $\Delta\varepsilon_{\text{Jagg}} = 0.075$, $\omega_{\text{abs}} = 2.9 \times 10^{15}$ s$^{-1}$ (649 nm) and $\gamma = 1.10 \times 10^{14}$ s$^{-1}$ ($\lambda_{\text{FWHM}} = 24$ nm) stand for the density, absorption frequency and line-width of the molecular exciton, respectively. These values were obtained from matching the extinction spectrum of J-aggregates coated PE layers with FDTD simulations. The side-length of the Au nanocubes in our simulations were set to be 65-nm with 8 corners replaced by 8-nm-radius spheres, where the permittivity Au were taken from the Ref. [8].

Specifically in simulations, the AuNC was illuminated by a plane-wave with electric field vertically oriented with respect to the film surface, as for all gap spacing in this study the optical properties of the NC-film cavity are dominated by the first order of vertically oscillating mode. (Detailed discussions can be found in the Section 4 of Supplementary Information) The gap spacing in our simulations for undoped cavities was adjusted to optimise the comparison with measured spectra, i.e. $d = 3$ nm for the 3 PE layer spacer. These spacing data were then applied in the modelling of doped cavities as well.

**Section. 2 Scattering Spectra and Patterns for NC-film cavities with various gap spacing**

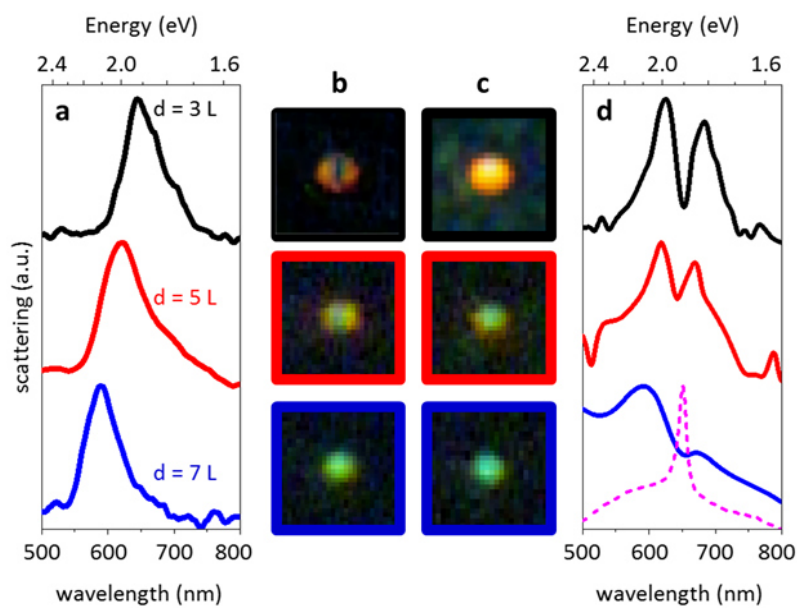

**Figure S1 Scattering properties of undoped and doped NC-film cavities.** Measured scattering spectra of undoped (a) and doped (d) NC-film cavities for gap spacing d = 3 (black curve), d = 5 (red) and d = 7 (blue) PE layers, with the corresponding far-field scattering images (b) and (c) respectively. The absorption spectrum (magenta dashed curve) of J-aggregates in aqueous solution (at a concentration of 10 µM dissolved with 1 mM sodium choloride) is shown in panel (d)

Fig. S1(a) shows typical scattering spectra of dye-free NC-film cavities with different gap spacing. Specifically, a nanocube placed 3 PE layers above the Au substrate gives rise to a maximum at 640 nm (1.94 eV) in the scattering spectrum (black curve in Fig. S1a), which blue-shifts with increasing gap spacing, such that the scattering spectrum shows a resonance peak at 610 nm for a gap of 5 PE layers and 588 nm for 7 PE layers gap spacing. The actual thicknesses of the PE spacers were measured using an ellipsometer and shown in the Table 1 of this Supplementary Information. The far-field scattering pattern also varies with the gap spacing. In particular, at the gap spacing of 3 PE layers, scattered light from the NC-film cavity shows a doughnut shape pattern (image with black frame in Fig. S1b), which gradually turns into a dot shape (blue framed in Fig S1b) with enlarged gap spacing up to 5 PE layers. These results concur with previous studies [9–11], indicating that due to the excitation of gap resonances, photons can be trapped within the gap between AuNCs and a metallic substrate.

Both the photon frequencies and far-field patterns are very sensitive to the gap geometry, as a result of different near-field distribution (both spectral and spatial) at varied gap spacing. Doping the nano-gap with J-aggregates leads to significant modifications in the optical properties. As shown in the main text, the doped NC-film cavity with a 3 PE layer gap (black curve in Fig. S1d) shows two pronounced split scattering maxima at 623 nm and 684 nm respectively with a deep minimum at 649 nm that shares the same spectral position of the absorption peak of the J-aggregates (magenta dashed curve in Fig. 1d). This is a typical feature of the occurrence of strong coupling between plasmonic modes and molecular excitons [12,13]. Increasing the gap spacing from 3 to 7 PE layers gradually diminishes such plexcitonic coupling features, e.g. with 7 PE layer spacing (blue curve in Fig. 2d), the scattering dip becomes very feeble. Comparing with scattering spectra of undoped cavities reveals that the best spectral overlap between plasmonic resonances (black curves in Fig. S1a) and molecular excitons (dashed magenta curve in Fig. S1d) gives the most pronounced splitting characteristics in the NC-film cavity with a 3 PE layer spacing (black curve in Fig S1d), further verifying the realization of a strong plexcitonic coupling scenario. Following Ref. [13], we note that the mode volumes $V$ of gap plasmons reduce as the gap spacing narrows, which also contributes to the coupling strength ($g$) and partly explains why the strongest plexcitonic coupling feature is observed in a cavity within the narrowest gap (3 PE layer spacing).

As shown in the main text, the spatial scattering profiles can also be significantly altered by the plexcitonic coupling: the scattering pattern of the doped NC-film cavity with 3 PE layer spacing (black framed in Fig. S1c) exhibits a dot shape, which is very different from the doughnut pattern shown in the undoped cavity (black framed in Fig. S1c).

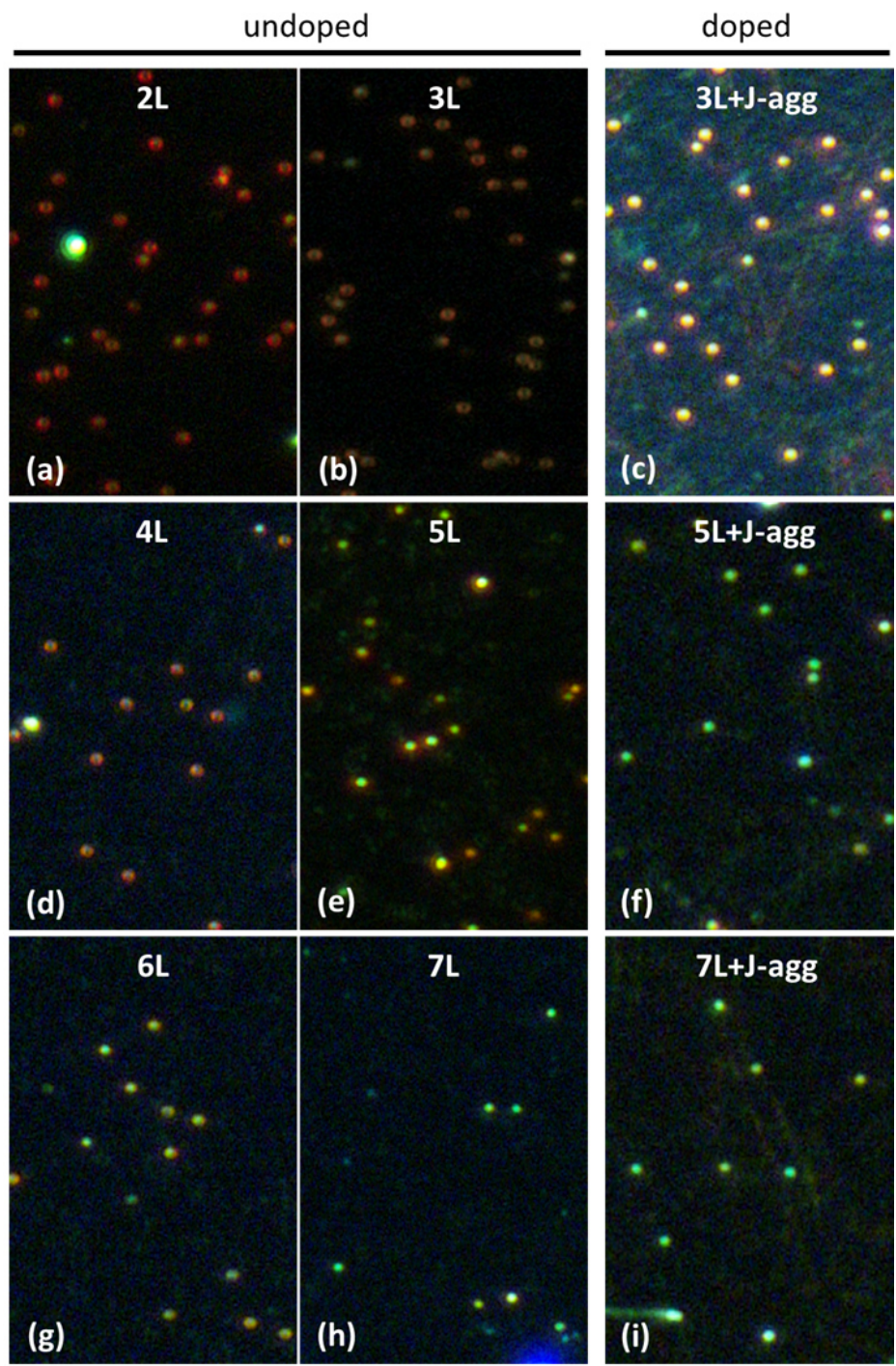

**Figure S2** Far-field scattering images for undoped cavities with different gap spacing: (a) 2 PE layers, (b) 3 PE layers, (d) 4 PE layers, (e) 5 PE layers, (g) 6 PE layers and (h) 7 PE layers; and doped NC-film cavities for (c) 3 PE layers+J-agg, (f) 5 PE layers+J-agg and (i) 7 PE layers+J-agg.

For gap spacing of 3 PE layers samples, the modification of scattering patterns is a systematic effect and not an artifact due to subtle variations in nanocube geometry or gap spacing. As shown in Fig S2, where the scattering images contain a number of NCs the in the same field of view for undoped and doped samples with various gap spacing, it is evident to see that in the same area, almost all NCs with gap spacing of 3 PE layers show doughnut shape scattering patterns (Fig. S2b), while almost all NCs with the same gap spacing show bright dot patterns (Fig. S2c). In contrast, the scattering images for doped cavities with 5 and 7 PE layer gap spacing (red and blue framed in Fig S1d respectively; Fig S2f and 2i respectively) all demonstrate patterns that are similar to those of their undoped counterparts (Fig. 2c) and Fig S2e and 2h, suggesting that this 'visible' modification of scattering patterns can only be observed when strong interactions between gap resonances and molecular excitons take place. This is likely because the frequency detuning between molecular excitons and gap resonances with larger gap spacing may hamper this effect as well.

**Section 3. Details of Sample Preparation**

**3.1 Structural configuration and J-aggregates**

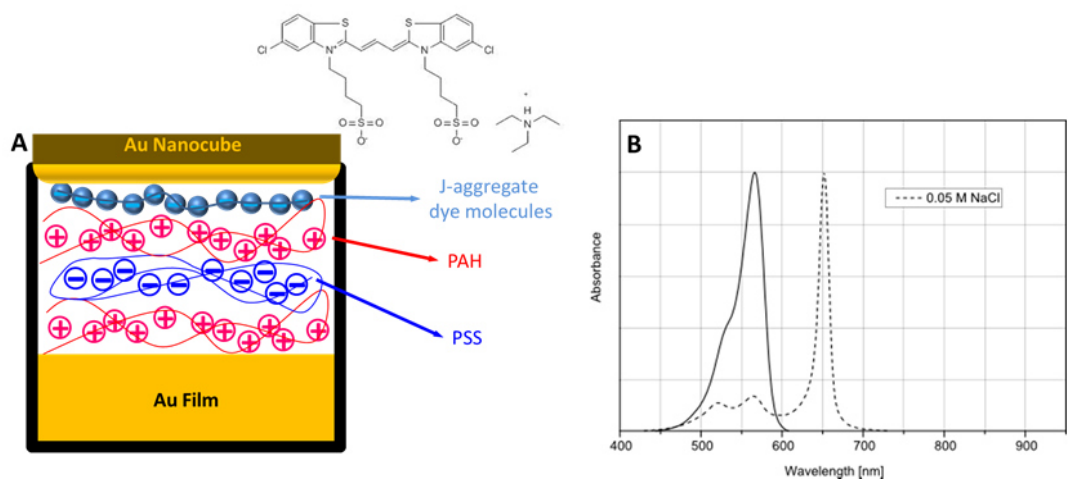

**Figure S3** (A) Schematic of NC-film cavity with a 3 PE layer spacer coated with J-aggregate dye molecules. Inset: structural formula of the dye monomer used in this study. (B) absorption spectra of the dye (used in this study) dissolved in methane (solid line) and water mixed with NaCl. (The panel B was obtained from the manufacturer)
http://www.few.de/en/functional-dyes/wasserloesliche-cyanine/500-699-nm/?type=3&tx_fewdyes_fewdyes%5Bfewdyes%5D=1282&tx_fewdyes_fewdyes%5Baction%5D=show&tx_fewdyes_fewdyes%5Bcontroller%5D=Fewdyes&cHash=207d053dad1f6ecbe4fc44f8fddc3898

Fig. S3A shows the schematic of a J-aggregates doped NC-film cavity with a 3 layer PE spacer. As mentioned in the main text, the cyanine dye molecules used in this study is "5-Chloro-2-[3-[5-chloro-3-(4-sulfobutyl)-3H-benzothiazol-2-ylidene]-propenyl]-3-(4-sulfobutyl)-benzothiazol-3-ium hydroxide, inner salt, triethyl ammonium salt" purchased from Few Chemicals, GmbH (Model No.: 2275). According to the structural formula (inset in Fig. S3A), the volume of the J-aggregate can be estimated as ~0.5 nm$^3$, which is at the same scale as the molecule used in the Ref. [13].

Fig. S3B shows the absorption spectra of the dye dissolved in methane (solid line), exhibiting an absorption peak at 566 nm. When dissolved in water (mixed with NaCl), the absorption maximum of the dye is highly narrowed and red-shifted with respect to the methane dissolved one. It means that in water solution the dyes are self-assembled from monomers [14] and become an aggregated form known as J-aggregates, as the resonance is significantly red-shifted but not blue-shifted (H-aggregates if blue-shifted). In the J-aggregate form, the long-axis of monomers (shown in the structural formula) is oriented almost parallel with the aggregation direction.

The J-aggregates are electrostatically attached to the surface of PE spacers. As mentioned in the main text, the dye molecules used in this study is negatively charged if dissolved in water, therefore the J-aggregate layer can only be adhered to positively charged PAH surfaces. According to the Ref. [15,16], if in an acidic environment, the electrostatic interaction within the PE layers is very strong, making it hard for dye molecules to penetrate into the spacers. Therefore the schematic shown in Fig S4A, in general, depicts the configuration of the J-aggregates coated PE spacer in a NC-film cavity.

### 3.2 Coverage of J-aggregates

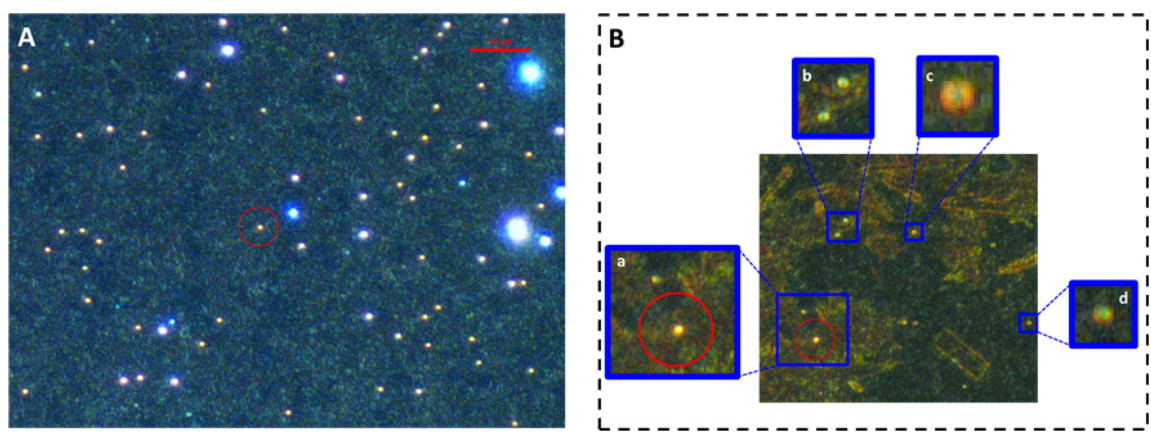

**Figure S4** Far-field scattering image for an area of 3 PE layer coated Au film, where J-aggregate layers are fully (**A**) and partially (**B**) covered. The different coverage areas are the result of different immersion time of PE-coated Au substrates in the dye solution. In panel **B**, nanocubes on the top are labelled and magnified with blue frames.

Principally, in our experiment, the J-aggregate molecules can fully cover most area of PE-layer coated Au substrates, so long as the PE spacer are terminated with PAH layer. However there are other factors that influence the coverage of J-aggregate molecules, e.g. the concentration of dye solutions, the immersion time in dye solutions and whether the dye solutions are mixed with HCl. Here we give a comparison between full and partial coverage of J-aggregate layers. Fig 4A shows the far-field scattering image of a PE-coated Au substrate fully and uniformly covered with J-aggregate molecules and then followed by AuNCs deposition. The immersion time of this sample in dye solutions is as standard as 10 min. It is clear to see that the background shows a uniformly distributed bluish-green color, while almost all AuNCs exhibit a dot shape scattering pattern. In contrast, if a PE-coated Au substrate was immersed in the dye solution shorter than 3min, the PE spacer surface is only partially covered, as shown in Fig. 3B. The background in this case exhibits patchy area with slightly yellowish color. More importantly, most of the AuNCs demonstrate doughnut-shape scattering patterns, whereas only 1 AuNC (Fig. S4Ba) show a dot-shape pattern, possibly due to the high concentration of dye molecules in its nearby area.

### 3.3 Thickness measurements

| Samples  Results | 3LBL | 3LBL+dye | 5LBL | 5LBL+dye | 7LBL | 7LBL+dye |
|---|---|---|---|---|---|---|
| Fitted thickness | 4.7 | 4.5 | 8.0 | 7.7 | 15.8 | 10.8 |
| Standard Deviation (1E-3) | 2.27 | 2.65 | 2.45 | 2.70 | 2.77 | 3.17 |
| | | | | | | |
| Fitted Thickness (nm) | 4.6 | 3.8 | 9.9 | 9.2 | 13.8 | 16.8 |
| Standard Deviation (1E-3) | 2.25 | 2.60 | 2.55 | 3.20 | 2.8 | 3.3 |
| | | | | | | |
| Fitted Thickness (nm) | 3.5 | 3.5 | 7.9 | 7.1 | 10.3 | 17.4 |
| Standard Deviation (1E-3) | 2.40 | 2.50 | 2.48 | 2.90 | 2.78 | 3.20 |
| | | | | | | |
| Averaged Thickness (nm) | 4.3±0.7 | 3.9±0.5 | 8.6±1.1 | 8.0±1.1 | 13.3±2.8 | 15.0±3.6 |

**Table 1**. Thickness of 3-, 5- and 7- PE layers uncoated or coated with J-aggregates

As mentioned in the main text, the thicknesses of PE spacers (uncoated and coated) were confirmed by ellipsometry measurements. Specifically, we have fabricated 3 samples for each set of spacers with specific number of PE layers (3, 5 and 7 layers in particular), and all PE spacers were prepared on Au substrates. The thicknesses of these samples were then measured using an ellipsometer (GES 5E Semilab). After measuring, these uncoated samples were then immersed in the dye solution and followed by being rinsed with distilled water to be covered with J-aggregates layer before being measured with the ellipsometer again. The ellipsometry measurement results were then fitted to obtain the thickness. For the fitting, refractive index of PE layers was set to be 1.45 given the site humidity and temperature [3,4], the index of J-aggregate coated PE layers was set using the value obtained from the formula in simulation (main text), and the index of gold substrate was obtained from the Ref. [8]. The net thickness of each set of samples was finally calculated as the mean value of the measurements as shown (red marked numbers) in Table 1. These results show that (i) as the number of PE layers increases, the thickness disproportionally expands. For example, in the case of 7 PE layers, the thickness has been increased up to $d = 13\pm3$ nm, which is not as thin as measured (~8 nm) in another study [4] probably due to the local humidity or temperature variation. (ii) The thicknesses of PE spacers coated with J-aggregates are generally thinner than the uncoated ones, especially for 3- and 5- PE layer spacers, possibly due to the pH-dependent swelling or de-swelling behaviors of PE layers as the dye solution has a pH value of $0.7\pm0.3$. Ref. [5] has reported that the thickness of PE layers shrinks in a low pH value environment, while in Ref. [17] the PE layers tend to expand in an acidic environment. However we note that the final thickness of PE layers significantly reduce after being repeatedly immersed in highly different pH environment [17], which may explain the shrunk PE spacers after being coated with J-aggregates, since the PE layers were immersed in acidic dye solution (pH=0.7) and then rinsed with de-ionized water (pH = 7).

The thicknesses of PE spacers terminated with PSS layers are shown in Table 2 as a reference.

| Results           Samples | 2LBL | 4LBL | 6LBL | 8LBL |
|---|---|---|---|---|
| **Fitted Thickness (nm)** | 2.2 | 5.8 | 10.6 | 15.9 |
| **Standard Deviation (1E-3)** | 2.16 | 2.33 | 2.55 | 2.76 |
| | | | | |
| **Fitted Thickness (nm)** | 4.5 | 7.0 | 11.2 | 15.3 |
| **Standard Deviation (1E-3)** | 2.26 | 2.40 | 2.61 | 2.74 |
| | | | | |
| **Fitted Thickness (nm)** | 2.9 | 5.1 | 10.4 | 14.1 |
| **Standard Deviation (1E-3)** | 2.24 | 2.50 | 2.59 | 2.70 |
| | | | | |
| **Measured Thickness (nm)** | **3.2±1.2** | **6.0±1.0** | **10.7±0.4** | **15.0±1.0** |

**Table 2**. Thickness of uncoated PE spacers for 2-, 4- and 8- PE layers.

**Section 4. Excitation**

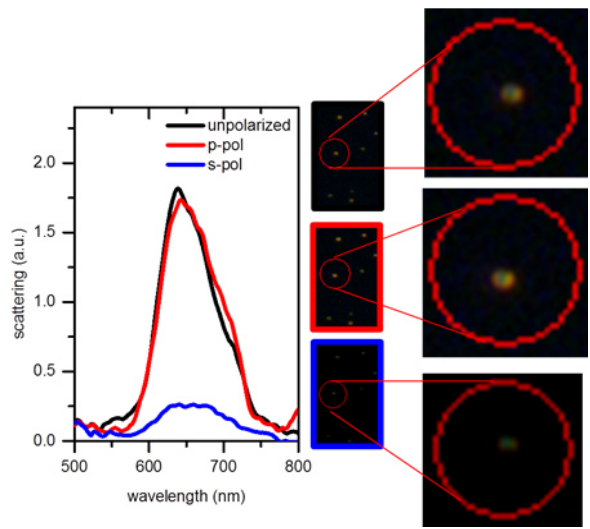

**Figure S5** scattering spectra of a single undoped NC-film cavity (3 PE layer spacer) excited with unpolarized (black solid curve), p-polarized (red solid curve) and s-polarized (blue solid curve) light; the corresponding far-field patterns are shown in the red circles of black (unpolarised), red (p-polarised) and blue (s-polarised) framed pictures.

The scattering spectra measured from the same undoped NC-film cavity (3 PE layer spacer) excited by white light with different polarizations are shown in Fig. S5. Illuminated by light with different polarizations, those spectra $I_{sc}$ were acquired as a function of wavelength using $I_{sc} = (S-R)/R$, where $S$ is the collected scattering signal from an area containing the single nanoparticle and $R$ is the signal acquired from a nearby area without the nanoparticle (the same sampling area). It turns out that the scattering spectra excited by unpolarized and p-polarized (electric field oriented parallel to the plane of incidence) light share similar spectral appearance and magnitude, while the spectrum excited by s-polarized (electric field oriented perpendicular to the plane of incidence) has much smaller magnitude. In addition, the far-field scattering patterns of NCs illuminated with unpolarised and p-polarised light all exhibit doughnut patterns, while the s-polarisation excited far-field patterns are very feeble. These indicate that the optical properties of the NC-film cavity with a 3 PE layer spacer are dominated by plasmonic resonances excited with p-polarized light.

**Seciton 5. Single PE layer spacer.**

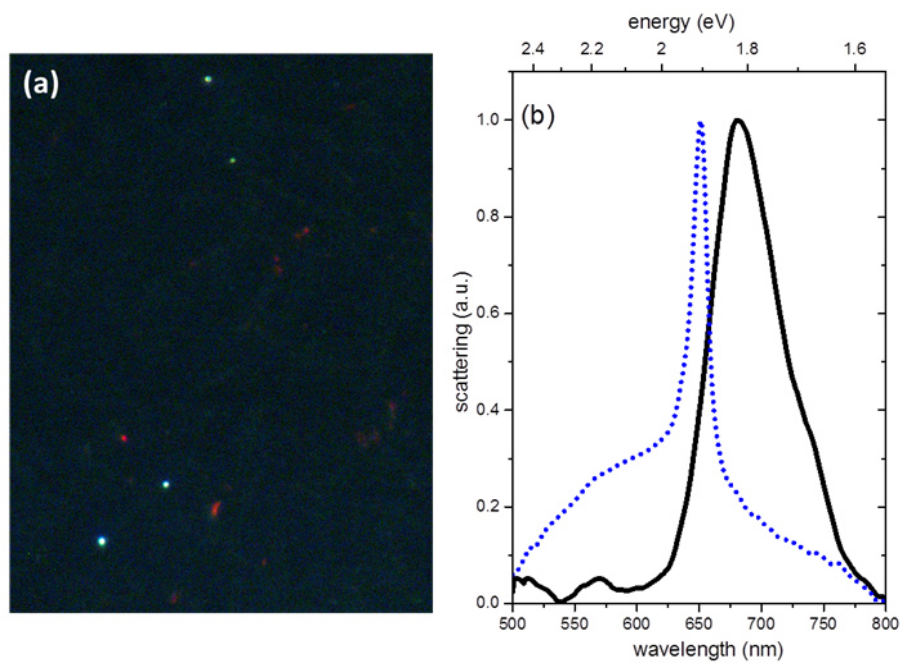

**Figure S6** (a) The scattering image of an area of J-aggregate coated single PAH layer on a Au substrate. (b) The measured spectrum of undoped NC-film cavities with 1 PE layer gap spacing.

As stated in the main text, using cavities with smaller gap spacing may reduce the gap mode volume, thus further enhancing the light-matter interaction in the nano-gap. To do so in our cavity system, we will have to deposit only one single PAH layer on the Au substrate to electrostatically adhere dye molecules. However, the surface of a single PAH layer can not be properly covered with J-aggregates layers as shown in Fig. S6(a) for unknown reasons.

In addition, the scattering spectrum of NC-film cavity with 1 PE layer spacer (black solid line) comparing with the absorption spectrum of J-aggregates (blue dashed line) are shown in Fig S6(b). The detuning between the dye absorption maximum (649 nm) and the gap resonance (~680 nm) may not lead to a pronounced strong plexcitonic coupling.